\input epsf.sty
\newdimen\psfigsize
\def\psfigure#1 #2 #3 #4 #5{
    \begin{figure}[tbh]
    \vbox{
    \null\vskip-0.2in\hskip#2\epsfxsize=#1 \epsfbox[0 0 4096 4096]{#4}
    \vskip -0.3in
    \caption {#5 \label{#3}}
    \vskip 0.0truein plus0.2truein}
    \end{figure}
}
\def\psoddfigure#1 #2 #3 #4 #5 #6{
    \begin{figure}[tbh]
    \vbox{
    \null\vskip-0.7in\hskip#3\epsfxsize=#1 \epsfbox[0 0 4096 4096]{#5}
    \vskip -#1 \vskip #2 \vskip 10truept
    \vskip -0.2in
    \caption {#6 \label{#4}}
    \vskip 0.0truein plus0.2truein}
    \end{figure}
}
\newcommand{\beq}{\begin{equation}}
\newcommand{\eeq}{\end{equation}}
\newcommand{\bea}{\begin{eqnarray}}
\newcommand{\eea}{\end{eqnarray}}
\newcommand{\bi}{\bibitem}

\documentstyle[twoside,fleqn,espcrc2]{article}

\title{Finite Temperature Lattice QCD with Clover Fermions
\thanks{Presented by Matthew~Wingate at Lattice 96.}
}

\author{ Claude~Bernard,\hskip-0.03in
\address{{\vskip-0.10in{\hskip 0.07in Department of Physics, Washington University, St.~Louis, MO 63130, USA}}} % "a"
Tom~Blum,\hskip-0.03in
\address{Department of Physics, Brookhaven National Lab, Upton, NY 11973, USA} %"b"
Thomas~A.~DeGrand,\hskip-0.03in
\address{Physics Department, University of Colorado, Boulder, CO 80309, USA} % "c"
Carleton~DeTar,\hskip-0.03in
\address{Physics Department, University of Utah, Salt Lake City, UT 84112, USA} % "d"
Steven~Gottlieb,\hskip-0.03in
\address{Department of Physics, Indiana University, Bloomington, IN 47405, USA} % "e"
Urs~M.~Heller,\hskip-0.03in
\address{SCRI, Florida State University, Tallahassee, FL 32306-4052, USA} %"f"
Jim~Hetrick,\hskip-0.03in
\address{Department of Physics, University of Arizona, Tucson, AZ 85721, USA} % "g"
Craig~McNeile,\hskip-0.03in$\,\null^{\rm d}$
Kari~Rummukainen,\hskip-0.03in$\,\null^{\rm e}$
Bob~Sugar,\hskip-0.03in
\address{Department of Physics, University of California, Santa Barbara, CA 93106, USA}
Doug~Toussaint$\,\null^{\rm g}$
and Matt~Wingate$\,\null^{\rm c}$
} %end \author
\begin{document}

\begin{abstract}

We report on our simulation of finite temperature lattice QCD
with two flavors of ${\cal O}(a)$ Symanzik-improved fermions
and ${\cal O}(a^2)$ Symanzik-improved glue.  Our
thermodynamic simulations were performed on an $8^3 \times 4$ lattice,
and we have performed complementary zero temperature simulations
on an $8^3 \times 16$ lattice.  We compare our results to those from
simulations with two flavors of Wilson fermions and discuss
the improvement resulting from use of the improved action.

\end{abstract}
\maketitle

%%%%%%
\section{INTRODUCTION}
\label{sec:intro}

The study of finite temperature QCD with Wilson-type quarks is
desirable in order to estimate any systematic errors of similar
simulations with Kogut-Susskind quarks.  However, Wilson thermodynamics
has proved to be difficult and burdened with lattice artifacts
\cite{ref:UKAWA}.  It is plausible that an action which converges
to the continuum action faster in the $a \rightarrow 0$ limit
would be cured of such spurious effects.

%%%%%%
\section{ACTION}
\label{sec:action}

For the gauge action, we start with the one loop, on-shell
Symanzik improved action derived by L\"uscher and Weisz \cite{ref:LW}.
We implement the tadpole improvement scheme in order that lattice
perturbation theory be more convergent \cite{ref:LM,ref:ALFORD}.
We choose to define the ``mean link'' $u_0$ and the strong coupling
constant $\alpha_s$ through the plaquette \cite{ref:LM,ref:ALFORD,ref:WW}:

The coefficients of the rectangle operator and the twisted 6-link operator, 
$\beta_{\rm rect}$ and $\beta_{\rm twist}$ respectively, are given in
terms of the coefficient of the plaquette $\beta$ and $u_0$ as in
\cite{ref:ALFORD}:
\bea
\beta_{\rm rect} & = & -{\beta \over 20 u_0^2}~\Big(1 - 0.6264~\ln(u_0)\Big) 
\label{eq:beta1} \\
\beta_{\rm twist} & = & {\beta \over u_0^2}~0.04335~\ln(u_0).
\label{eq:beta2}
\eea
In practice, we estimate $u_0$ in a self-consistent manner:  we
tune it so that it agrees with the fourth root of the space-like
plaquettes.

The Wilson fermion action has errors of ${\cal O}(a)$.  The
Symanzik improvement program is used to improve the action
\cite{ref:SW}. After tadpole improvement the fermion action is
\beq
S_f ~=~ S_W - {\kappa \over u_0^3} \sum_x \sum_{\mu < \nu} \bigg[ 
\overline{\psi}(x) ~ i \sigma_{\mu\nu} F_{\mu\nu} \psi(x) \bigg],
\eeq
where $S_W$ is the usual Wilson fermion action, and $i F_{\mu\nu}$ is
the familiar clover-shaped link operator.

\psoddfigure 3.0in 2.7in -0.2in {fig:phase} {phase_cl_lat96.ps} {
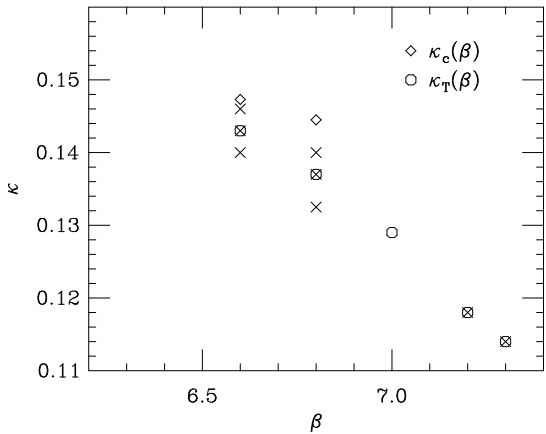
Phase diagram of Symanzik-improved action.  Octagons represent
the $N_t=4$ thermal crossover, and diamonds indicate estimates of vanishing
pion mass.  Zero temperature simulations were performed at the crosses.
}

\psoddfigure 3.0in 2.7in -0.2in {fig:pl_vs_k} {PL_vs_k_lat96.ps} {
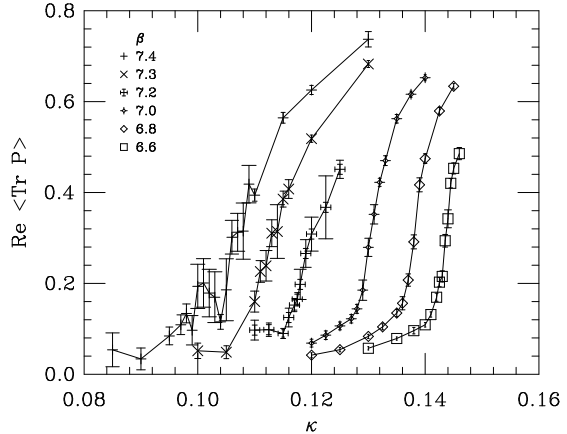
Polyakov loop vs. hopping parameter for $8^3 \times 4$ improved
Wilson thermodynamics.
}

%%%%%%
\section{RESULTS}
\label{sec:results}

Our thermodynamics simulations were done on an $8^3 \times 4$ lattice
at six fixed values of $\beta$ while varying $\kappa$ across
the thermal crossover (tuning $u_0$ self-consistently at each parameter
set).  We used the hybrid Monte Carlo algorithm and collected data from
at least 1000 trajectories for the simulations in the crossover region.
Furthermore, zero temperature simulations on an $8^3 \times 16$ lattice
were performed in order to provide hadron masses in the region of the thermal 
crossover line.  The phase diagram (figure~\ref{fig:phase}) summarizes
our run parameters.

Figure~\ref{fig:pl_vs_k} shows the Polyakov loop as a function of $\kappa$
for the six values of $\beta$.  One can observe that the transition
appears steeper for stronger coupling: a feature also present
in $N_t=4$ Wilson thermodynamics~\cite{ref:MILC4}.
Still, the crossover for the improved action does not appear to be
as steep as for the unimproved action.

One would like to make direct comparison 
between the two actions of their respective crossover behavior
without depending on the bare parameters.  In this work, we use 
measurements of the lattice pion mass squared at values of $\kappa$
near the crossover.  Then, we can plausibly overlay curves of thermodynamic
observables for two actions run at comparable $m_\pi/m_\rho$.  Below we
list $m_\pi/m_\rho$ along the $N_t=4$ crossover for both clover and
Wilson \cite{ref:HEMCGC} actions.
%\begin{table}
%\caption{\label{tab:spect} $m_\pi/m_\rho$ along the $N_t=4$ thermal crossover.}
\begin{center}
\begin{tabular}{cc|c||c|cc}
\multicolumn{3}{c}{Clover} & \multicolumn{3}{c}{Wilson} \\ \hline
$\beta_{\rm Cl}$ & $\kappa_{\rm Cl}$ & $m_\pi/m_\rho$
& $m_\pi/m_\rho$ & $\beta_{\rm W}$ & $\kappa_{\rm W}$ \\ \hline
6.6  &  0.143  &  0.725(24)  &  0.708(7)  &  4.76  &  0.19 \\
6.8  &  0.137  &  0.831(10)  &  0.836(5)  &  4.94  &  0.18 \\
7.2  &  0.118  &  0.968(4)   &  0.899(4)  &  5.12  &  0.17 \\
7.3  &  0.114  &  0.970(3)   &  0.943(5)  &  5.28  &  0.16
\end{tabular}
\end{center}
%\end{table}

\psfigure 3.0in -0.2in {fig:pl_vs_mpisq} {rp_vs_mpisq2_lat96.ps} {
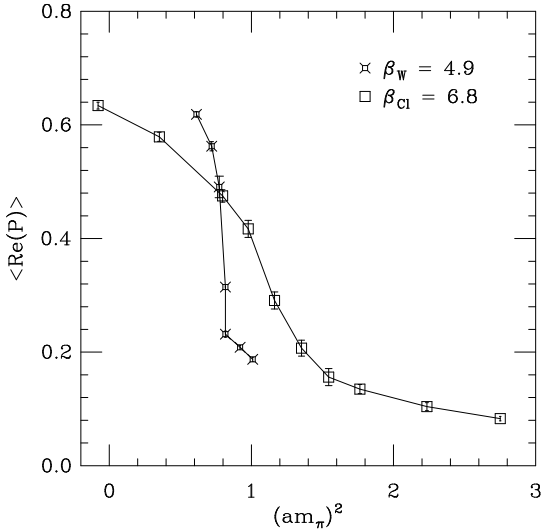
Polyakov loop vs.\ pion mass squared.  The crossovers ($N_t=4$) for both
actions, at the couplings shown, occur at the same $\pi-\rho$
mass ratio: $m_\pi/m_\rho = 0.83$.
}

Using measurements of the pion mass near the crossover region \cite{ref:SCRI},
we can interpolate in order to estimate $(am_\pi)^2$ as a function of
$1/\kappa$.  Then, we can plot the thermodynamic observables against
the pion mass squared.  This shows that the crossover is indeed
smoother for $N_t=4$ clover than $N_t=4$ Wilson
(see figure~\ref{fig:pl_vs_mpisq}).

\psfigure 3.0in -0.2in {fig:tc_over_mrho} {tc_over_mrho_lat96.ps} {
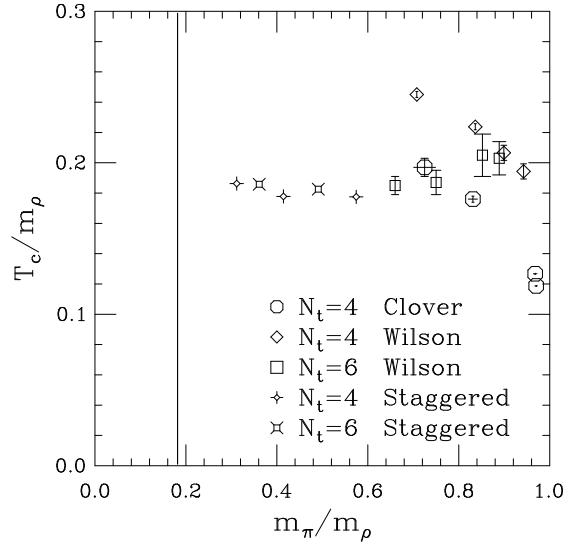
Crossover temperature in units of the $\rho$ mass vs.\  
$m_\pi/m_\rho$.
}

Finally, the confinement-deconfinement temperatures for
different two-flavor lattice actions are shown in
figure~\ref{fig:tc_over_mrho}.
One consequence of our improvement scheme is to lower
the Wilson $N_t=4$ critical temperature at a given mass ratio.
This brings the calculation of $T_c/m_\rho$ into better agreement
with $N_t=6$ Wilson and with staggered fermion thermodynamics.
There is one important caveat:  we do not yet know if the clover
simulations have reached a plateau in $m_\pi/m_\rho$.  If
$T_c/m_\rho$ continues to rise at lower $m_\pi/m_\rho$
(lower $\beta$) then the aforementioned agreement is accidental.
Furthermore, we should remember that $T_c/m_\rho$ tends toward
zero as $m_\pi/m_\rho$ approaches unity, {\it i.e.}\ as
$m_q \rightarrow \infty$.  Measurements of the string tension
will provide a scale which is insensitive to the quark mass.

\section{CONCLUSIONS}
\label{sec:concl}

We have shown that the $N_t = 4$ thermal crossover is smoother
for the Symanzik-improved action.  It could be that $T_c/m_\rho$
is in better agreement with staggered fermion results; however
running at the thermal crossover at lower $m_\pi/m_\rho$ is
needed to confirm this.

This work was supported by the U.S.\ Department of Energy and
the National Science Foundation.

%%%%%%%%%%%%%%%%%%%%%%%%%%%%%%%%%%%%%%%%%%%%%%%%%%%%%%%%%%%%%%%%

%%%%%%%%%%%%%%%%%%%%%%%%%%%%%%%%%%%%%%%%%%%%%%%%%%%%%%%%%%%%%%%%

\end{document}